\documentclass[preprint2]{aastex}
\usepackage{color}

\newcommand{\teff}{$T_{e\!f\!f}$}
\newcommand{\kms}{km s$^{-1}$}
\newcommand{\logg}{$\log g$} 

\slugcomment{Not to appear in Nonlearned J., 45.}
\usepackage{epsfig}

\shorttitle{Split clump}
\shortauthors{M Ness et al. }

\begin{document}


\title{THE ORIGIN OF THE SPLIT RED CLUMP IN THE GALACTIC BULGE OF THE MILKY WAY}


\author{M. NESS\altaffilmark{1}, K. FREEMAN\altaffilmark{1}, E. ATHANASSOULA\altaffilmark{2}, E. WYLIE-DE-BOER\altaffilmark{1}, J. BLAND-HAWTHORN\altaffilmark{3}, G.F. LEWIS\altaffilmark{3}, D. YONG\altaffilmark{1}, M. ASPLUND\altaffilmark{1}, R. R. LANE\altaffilmark{4}, L. L. KISS\altaffilmark{3}$^{,}$\altaffilmark{5}, R. IBATA\altaffilmark{6}}
\affil{$^1$Research School of Astronomy \& Astrophysics, Australian National University, Cotter Rd., Weston, ACT 2611, Australia}
\affil{$^2$LAM, UAM \& CNRS, UMR7326, 38 rue F. Joliot-Curie, Marseille 13, France.} 
\affil{$^3$Sydney Institute for Astronomy, University of Sydney, School of Physics A28, NSW 2006, Australia} 
\affil{$^4$Departamento de Astronom\'{i}a Universidad de Concepci\'{o}n, Casilla 160 C, Concepci\'{o}n, Chile} 
\affil{$^5$Konkoly Observatory, MTA Research Centre for Astronomy and Earth Sciences, Budapest, Hungary} 
\affil{$^6$Observatoire Astronomique, Universit\'{e} de Strasbourg, CNRS, 11 rue de l'Universit\'{e}, F-67000 Strasbourg, France } 
 \email{mkness@mso.anu.edu.au}


\begin{abstract}

Near the minor axis of the Galactic bulge, at latitudes $b <-5^\circ$, the red giant clump stars are split into two components along the line of sight. We investigate this split using the three fields from the ARGOS survey that lie on the minor axis at $(l,b) = (0^\circ,-5^\circ), (0^\circ,-7.5^\circ), (0^\circ,-10^\circ)$. The separation is evident for stars with [Fe/H] $> -0.5$ in the two higher-latitude fields, but not in the field at $b = -5^\circ$. Stars with [Fe/H] $< -0.5$ do not show the split.  We compare the spatial distribution and kinematics of the clump stars with predictions from an evolutionary N-body model of a bulge that grew from a disk via bar-related instabilities.  The density distribution of the peanut-shaped model is depressed near its minor axis. This produces a bimodal distribution of stars along the line of sight through the bulge near its minor axis, very much as seen in our observations. The observed and modelled kinematics of the two groups of stars are also similar. We conclude that the split red clump of the bulge is probably a generic feature of boxy/peanut bulges that grew from disks, and that the disk from which the bulge grew had relatively few stars with [Fe/H] $< -0.5$.
\end{abstract}


\section{INTRODUCTION}

The Milky Way has a boxy/peanut-shaped bulge, as seen in the 2MASS star 
counts (L{\'o}pez-Corredoira et al. 2005) and the COBE/DIRBE near-infrared light
distribution (Dwek et al. 1995). The long axis of this bulge/bar 
lies in the Galactic plane and points into the first Galactic
quadrant at an angle of about $20^\circ$ to the Sun-center line
(Gerhard 2002). The distribution of stars associated with
the boxy/peanut structure shows complexity at higher Galactic latitudes.
Near the minor axis, at $b < -5^\circ$, recent analyses of optical 
and near-IR stellar photometry by Saito et al. (2011), Nataf et al. (2010) and McWilliam \& Zoccali (2010) show that the red giant clump stars are split 
into two components separated by about 0.65 mag along the line of 
sight. McWilliam \& Zoccali (2010) argue that the split clump is
associated with an underlying X-shaped structure in the bulge. Such 
X-shaped structure has been noted earlier for other boxy/peanut
bulges: (e.g. Whitmore \& Bell (1988)).

We have recently completed the Abundances and Radial velocity Galactic 
Origins Survey (ARGOS) which is a spectroscopic survey of 28,000
stars, mostly clump giants, in 28 fields of the Galactic bulge and nearby 
disk.  The ARGOS survey provides a clean sample of bulge giants with distances, 
radial velocities and metallicities.  As part of this survey, we have data 
for about 3000 stars in three two-degree fields along the minor axis, at 
$(l,b)$ = $(0^\circ,-5^\circ), (0^\circ,-7.5^\circ)$ and $(0^\circ,-10^\circ)$. 
These fields are hereafter denoted as m0m5, m0m75 and m0m10. We find an
obvious split in the magnitude distribution of the red clump stars along the 
minor axis in the higher-latitude fields, m0m75 and m0m10.

 The objectives of this paper are (1) to measure the metallicity range of the stars that show the split red clump near the minor axis of the bulge, (2) to determine whether the split red clump has an associated kinematic signature, and (3) to interpret these results in the context of an evolutionary N-body model.  Our study complements previous photometric analyses of the stellar density and magnitude distribution and expands upon previous spectroscopic studies: e.g Rangwala \& Williams (2009); De Propis (2011).

\section{OBSERVATIONS}

\begin{figure}
\includegraphics[scale=0.27]{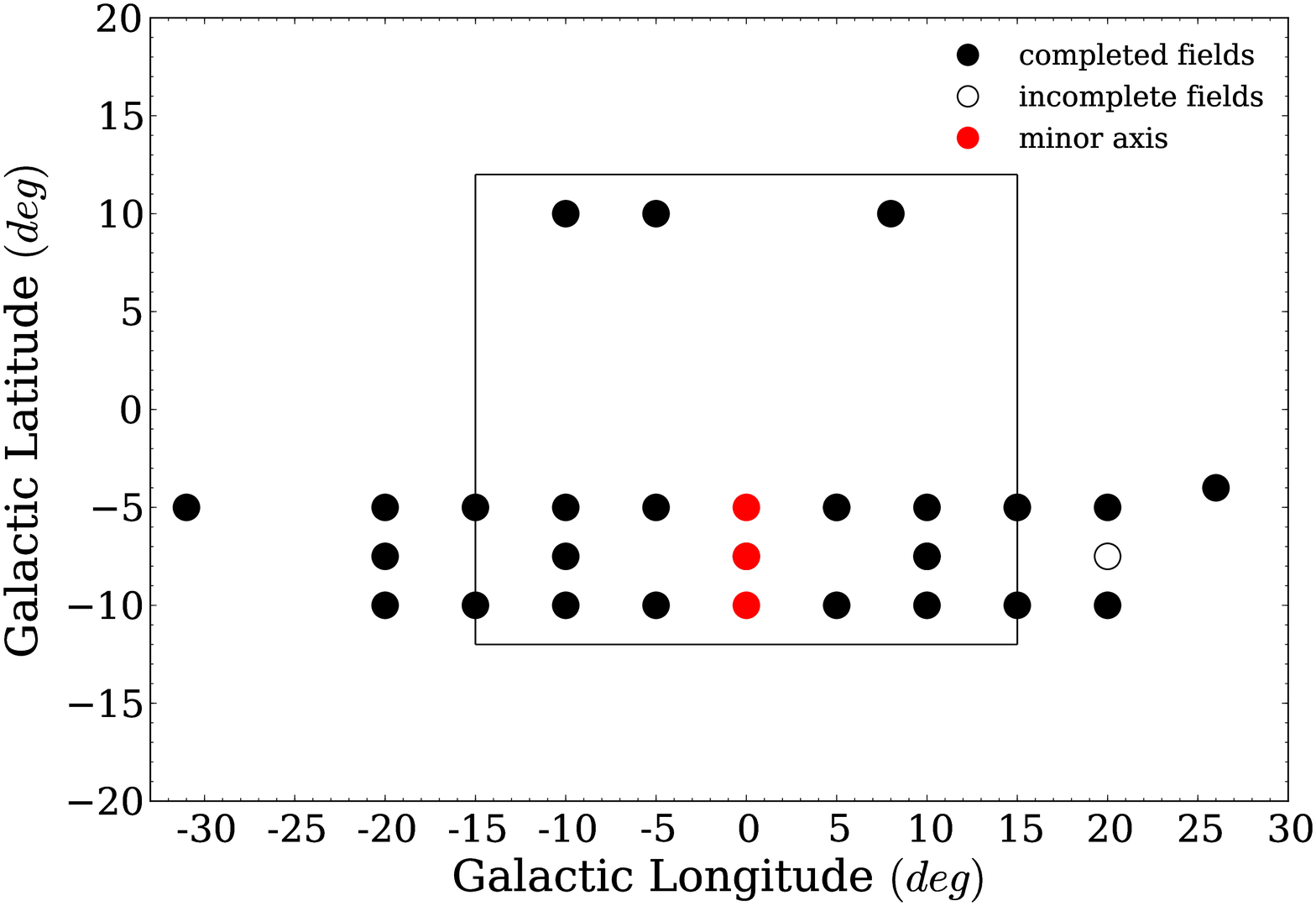}
\caption{The circles show the location of the ARGOS fields. The rectangle roughly 
indicates the outer limits of the Galactic bulge. The three minor axis fields are 
shown in red. The survey is now complete, except for the field at $(l,b)$ = 
$(20^\circ,-7.5^\circ)$ for which only 600 stars were observed. }
\label{fig:fields}
\end{figure}

\begin{figure}
\includegraphics[scale=0.38]{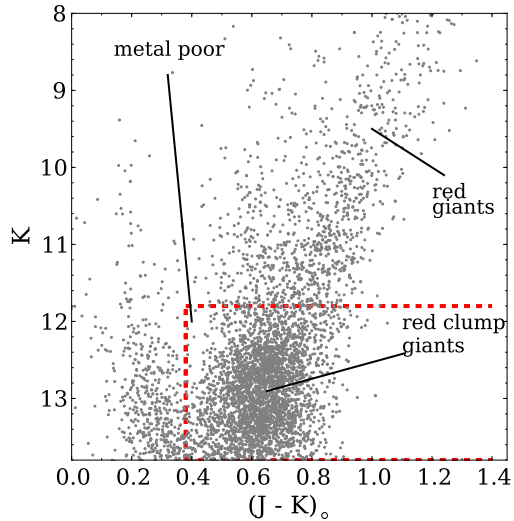}
\caption{The $K,(J-K)$ CMD for stars in the ARGOS m0m75 field at 
$(l,b) = (0^\circ, -7.5^\circ)$.  Stars for observation were selected from 
within the area outlined by the dashed box, which includes bulge giants 
with a wide range of metallicity and excludes most of the foreground disk 
dwarfs. The red clump giants lie near $(J-K)_\circ$ = 0.65.}
\label{fig:cmd75}
\end{figure}

The ARGOS observations were taken during 42 nights in 2008-2011 with the Anglo 
Australian Telescope at Siding Spring Observatory. We used the AAOmega fibre-fed 
spectrograph (Sharp et al. 2006) to observe about 340 stars simultaneously. The 
spectral region was $8400-8800$ \AA\ at a resolution of about 11,000. The locations 
of the 28 fields for the survey are shown in Figure \ref{fig:fields}. About 1000 
stars were observed in each field: the spectra have a typical signal to noise of 
50 to 80 per resolution element. We selected stars from the Two Micron All Sky 
Survey (2MASS) in the magnitude range $K = 11.5$ to $14$ with errors in $J,K < 0.06$ 
and with 2MASS flags set to reduce blends, contamination and stars of lower
photometric quality. The colour-magnitude diagram (CMD) for our m0m75 field  
in Figure \ref{fig:cmd75} shows our colour and magnitude selection criteria. 

To make our selection cuts, we first dereddened the stars using the mean 
Schlegel et al. (1998) reddening for each field. For the three minor axis fields,
the mean reddening is $E(J-K) = 0.37, 0.15$ and $0.09$ for our m0m5, m0m75 and 
m0m10 fields respectively. Because the primary observations were made at the
Ca triplet, and we wanted to observe stars of similar brightness together,
we estimated their $I$-magnitudes using a simple transformation from $J-K$ to 
$I-K$ derived from Bessell \& Brett (1988). For  the fibre allocations we then
selected about 340 stars in each one-magnitude interval of $I$, from about
$13$ to $16$.  The integration times for these three magnitude intervals were $60$ 
minutes, $120$ minutes and $150$ minutes, respectively. Full details of the ARGOS 
survey will be presented in Freeman et al. (2012, in preparation). 

\section{DATA REDUCTION, RADIAL VELOCITIES AND STELLAR PARAMETERS}

We reduced our data using the standard 2dfdr pipeline\footnote{http:
//www2.aao.gov.au/twiki/bin/view/Main/CookBook2dfdr}. This
program subtracts bias and scattered light, extracts the individual
stellar spectra, corrects for the flat field, calibrates wavelength
and fibre throughput and subtracts sky.  We selected the optimal
sky subtraction option which gives much better sky subtraction, and
used a 4th order polynomial for fitting the calibration lamp lines.

We determined radial velocities and \teff, \logg, [Fe/H], and [$\alpha$/Fe]
for all stars, using our ARGOS stellar pipeline (Ness et al. 2012, in preparation), and estimated their 
distances via the Basti isochrones, assuming an age of 10 Gyrs. 

Radial velocities were obtained via cross correlation of the
observed stellar spectra with synthetic spectra. The template spectra were generated using the local thermodynamic equilibrium (LTE) stellar synthesis program MOOG (Sneden 1973) using 1D LTE atmospheres (Casteli \& Kurucz 2003).

The stellar parameters of the template spectra covered a range of
[Fe/H] at a fixed gravity and temperature. The metallicity values
used were [Fe/H] = $-$2.0, $-$1.0 and 0.0, for \teff \ = 5000 K and
$\log g = 2.8$ appropriate for a typical clump giant (Zhao et al. 2001).
The best matching template to each observed star, as determined by the minimum error analysis via cross correlation,  was chosen to
determine the radial velocity.  The radial velocity errors at our observed resolution
(11,000) and signal-to-noise ($\sim 50$ to $80$ per resolution
element) is $< 1.2$ km s$^{-1}$.

The heliocentric radial velocities were transformed to Galactocentric 
velocities. We took the circular velocity of the local standard of rest 
to be $220$ \kms\ (Kerr \& Lynden-Bell 1986) and the  peculiar velocity of the sun to 
be $16.5$ \kms\ in the direction $(l,b) = (53^\circ,25^\circ)$.

Stellar parameters were determined using a minimum $\chi^2$ program
written for this dataset; it compares each observed spectrum to a
large grid of synthetic spectra. The grid of synthetic spectra was
generated using the LTE stellar synthesis program MOOG 2010\footnote{www.as.utexas.edu/$\sim$chris/moog.html} from the Castelli \& Kurucz (2003) model 
atmospheres with no convective overshooting. The final grid covers a 
parameter range of \teff \ $ =  3750 - 6250$ $K$, [Fe/H] $ = -5.0$ to $0.0,       
\log g  = 0.0$ to $5.0$ and [$\alpha$/Fe] $ = 0.0$ to $+1.0$. The grid steps are $250$ K 
in \teff, $0.5$ in $\log g$, $0.5$ in [Fe/H] (i.e. the steps of the original opacity distribution functions) and 0.1 dex in [$\alpha$/Fe], for a microturbulent velocity of  $2.0$ \kms. 

The $\chi^2$ program was verified and tested using a number of
calibration stars and fields and the Solar and Arcturus spectra
from the Hinkle Atlas (Hinkle et al. 2000). Full details of 
the $\chi^2$ program, including our linelist, grid, implementation and calibration will  be given in a forthcoming paper by Ness et al. (2012, in preparation). 

Using our spectroscopic stellar parameters, we are able to select a clean sample 
of genuine clump giants. This sample can be further refined by using distances for
our stars estimated from the derived stellar parameters. The apparent $K$ magnitude of 
each star was corrected for interstellar extinction,
using the Schlegel reddening at the position of each individual
star and adopting $A_K = 0.53 E(B-V)$ (Schlegel et al. 1998). The correction is typically 
$A_{K} = 0.25, 0.1$ and 0.05 for our three minor axis fields m0m5,
m0m75 and m0m10 respectively. For the clump giants (identified
as such from their temperatures and gravities) we adopted an absolute
magnitude for the clump of $M_{K} = -1.61$ (Alves 2000).  For
stars not in the clump, distances were estimated via fits to the
Basti isochrones (Cassisi et al. 2006), adopting an age of 10 Gyr. For the clump stars,
the distance errors come from the uncertainties in (i) the interstellar
reddening, (ii) the clump absolute magnitude and (iii) 
the 2MASS apparent magnitude. We
estimate our distance errors to be $< 1.5$ kpc.  These distance
estimates allow us to exclude stars which are not in the region of
the bulge.

In the following sections, we will discuss the magnitude distribution
and kinematics of the split red clump seen in the bulge near its
minor axis.

\section{THE ORIGIN OF THE SPLIT RED CLUMP}

McWilliam \& Zoccali (2010) found two red clump populations in 
the bulge at latitudes $|b| > 5.5^\circ$, separated by about 0.65
mag ($K_\circ = 12.64$ and $13.29$) or $2.3$ kpc along the line of sight in their field near the minor
axis $(l,b)=(1^\circ,-7.5^\circ)$. 

From the three ARGOS survey fields on the minor axis of the bulge,
we have a sample of confirmed red clump stars with known velocities,
distances and metallicities.  First we would like to determine the
metallicity range over which the split red clump is seen. The two higher-latitude
fields m0m75 and m0m10 analysed separately show very similar results and to increase the number of stars in our analysis, they are combined for this discussion: the stars in these
fields have $b$ between $-6.5$ and $-11^\circ$ and $l$ between $\pm
1^\circ$. To ensure a sample of clump stars lying in the bulge
region, we select stars with $\log g = 1.9$ to $3.1$ and $K_\circ
= 12.38$ to $13.48$. 

\begin{figure}[!h]
\centering
\includegraphics[scale=0.25]{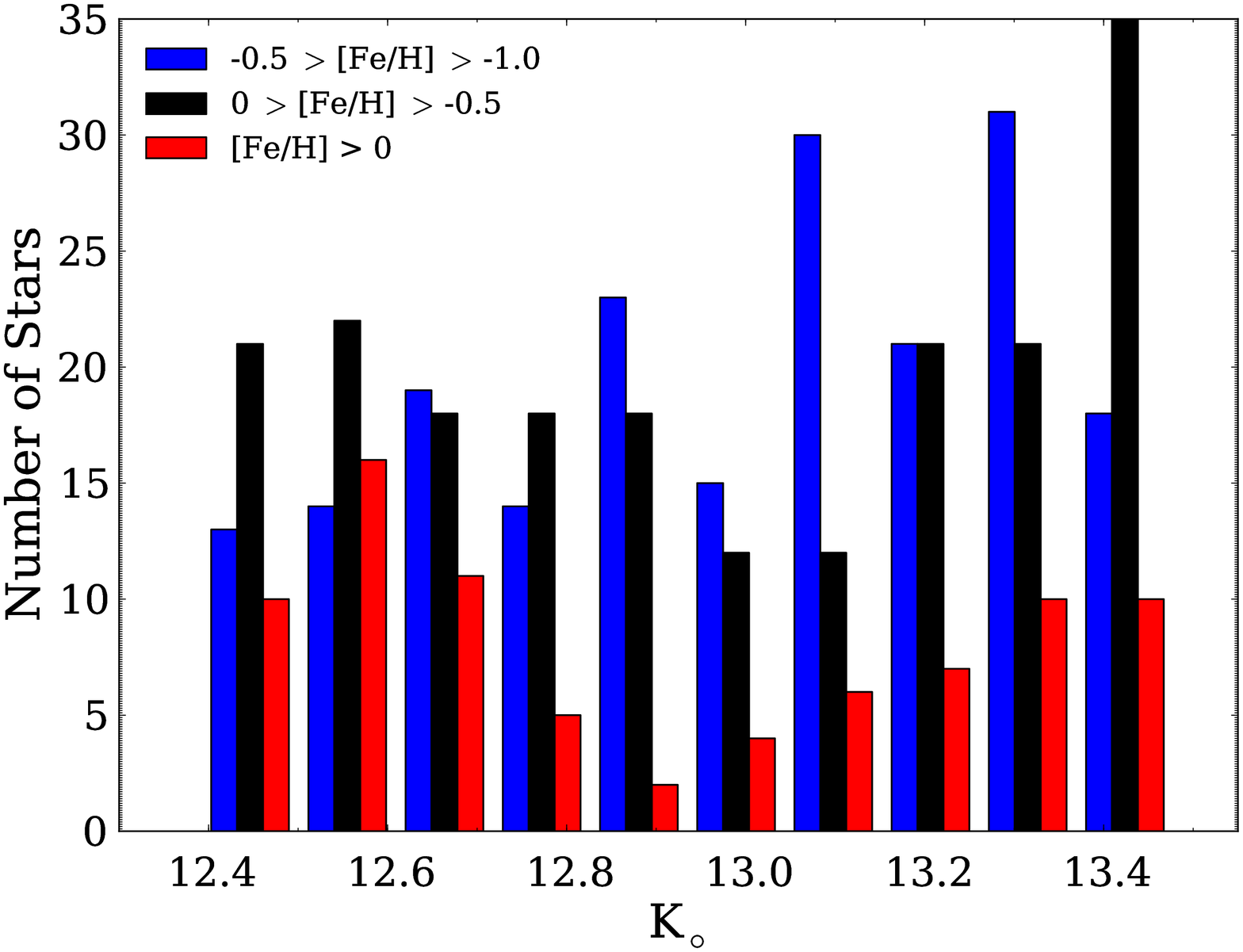}
\includegraphics[scale=0.25]{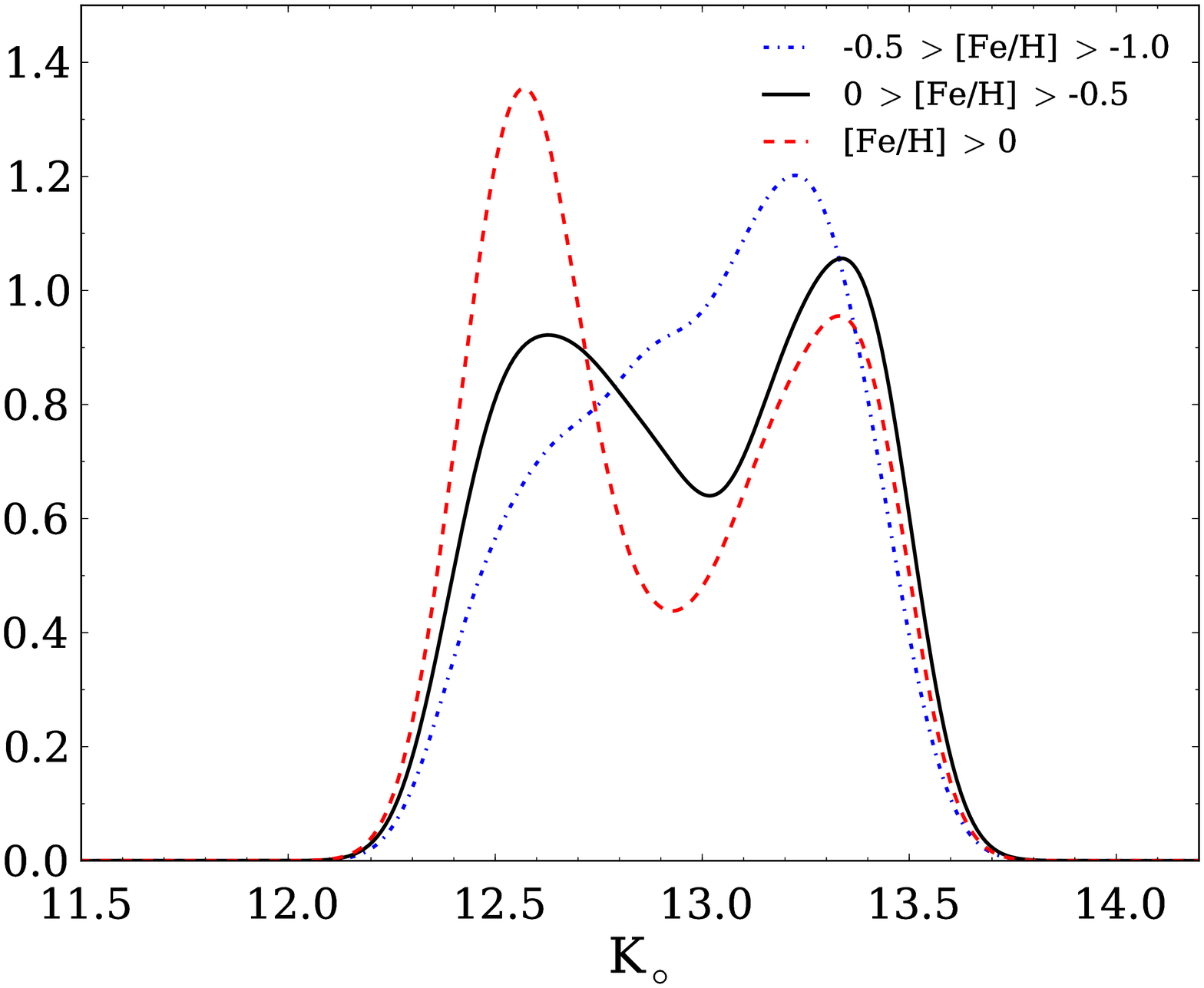}
\caption{Magnitude distributions for the clump stars in m0m75 + m0m10 selected with $12.38 <$ $K_\circ$ $< 13.48$, 1.9 $<$ log g $<$ 3.1 and (i) Fe/H $>$ 0 (81 stars) (ii) 0 $>$ Fe/H $>$ --0.5 (190 stars), (iii) --1.0 $>$ Fe/H $>$ --0.5 (206 stars). The bimodal distribution is seen only for stars with [Fe/H] $>$ --0.5 and is most prominent for stars [Fe/H] $>$ 0. The top figure shows the binned data across the K magnitude range for each metallicity interval, and the bottom figure shows the normalised generalised histograms (kernel $\sigma$ = 0.1) which show the minimum in the K magnitude distribution for stars [Fe/H] $>$ --0.5. The vertical axis units for the bottom figure are arbitrary.}
\label{fig:xshape}
\end{figure}

Examination of the K magnitude distribution of the clump stars as a 
function of [Fe/H] shows that the bimodality in the red clump magnitude distribution
is very clear for the 271 red clump stars with [Fe/H] $> -0.5$ in 
the higher-latitude fields; it does not appear to be present for the 206 
more metal-poor stars with $-0.5 \ge $ [Fe/H] $> -1.0$. Figure \ref{fig:xshape} compares the magnitude distributions for stars in the m0m75 + m0m10 fields across three [Fe/H] intervals: (i) [Fe/H] $> 0$, (ii) $0 >$ [Fe/H] $> -0.5$ and (iii) $-0.5 > $ [Fe/H] $> -1.0$. 
The binned data are shown in the upper panel and the normalised generalised distributions with Gaussian kernel $\sigma$ = 0.10 mag in the lower panel. In the generalised histograms, each star is represented by a gaussian distribution with mean at the magnitude of the star and $\sigma$ = 0.10 mag (the maximum likely error of the K magnitudes for our stars). We use generalised histograms to avoid the effects of arbitrary bin  boundaries in conventional histograms. The generalised histograms clearly demonstrate the structure of the distribution.
We do not show the stars with [Fe/H] $< -1.0$ here, because they probably belong to the inner metal-poor thick disk and halo (Ness et al., 2012, in preparation). The peak of the magnitude distribution for stars with $-0.5 >$ [Fe/H] $> -1.0$ lies at a fainter K magnitude than the mean of the more metal-rich stars, probably because we preferentially sample more of the more distant metal-poor stars in our survey cone at higher latitudes. The magnitude distribution of the stars with [Fe/H] $< -1.0$ is 15\% narrower than for stars with $-1.0 >$ [Fe/H] $> -0.5$ and, similarly, does not show the split clump.

We note that in the higher-latitude fields the bimodality is seen very clearly for the most metal-rich red clump stars with [Fe/H] $> 0$, although they are kinematically different from the rest of the bulge. Their velocity dispersion is somewhat smaller ($60 \pm 5$ \kms) than for the more metal-poor bulge stars with [Fe/H] $= -0.5$ to $0.0$ ($81 \pm 3$ \kms) which are the dominant population in the bulge. The split for the stars with [Fe/H] $>$ 0 is marginally larger than for the stars with $0.0 >$ [Fe/H] $> -0.5$, and we sample more of the brighter metal rich stars on the near side of the bulge because they lie preferentially at lower latitudes. We will return to this point in our discussion of the metallicity components of the bulge (Ness et al. 2012, in preparation).

\begin{figure}[!h]
\centering
\includegraphics[scale=0.3]{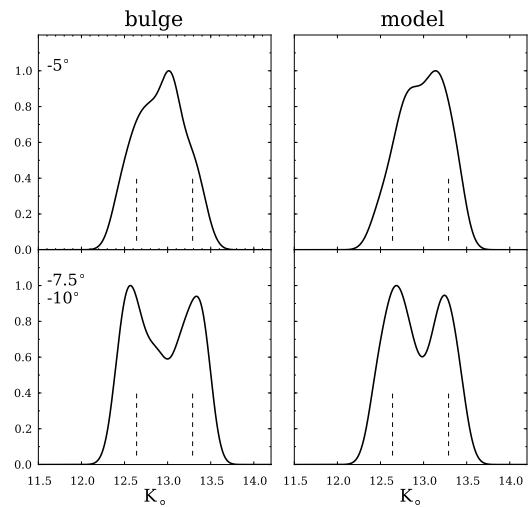} 
\caption{The K-band magnitude distributions of clump stars red with [Fe/H] $>$ --0.5 in the ARGOS
m0m5 field ({\it upper left}) and in the two combined fields m0m75 + m0m10
({\it lower left}). Only stars with $12.38 <$ $K_\circ$ $< 13.48$ are included.
The magnitude distributions are shown as generalised histograms, using
a Gaussian kernel with $\sigma = 0.10$ mag.  The panels on the right
show the predicted magnitude distribution for the N-body model, adopting $M_{K}$ = --1.51 for the clump stars. The vertical dashed lines show
the magnitudes of the two groups reported by McWilliam \& Zoccali (2010)
in their bulge fields near the minor axis. The bimodal magnitude distribution is seen clearly for the bulge and the model in the higher 
latitude fields but only weakly in the m0m5 field. The vertical axis units are arbitrary.}
\label{fig:fig4}
\end{figure}

We now examine the latitude dependence of the bimodality for the red clump stars which show the split red clump (those with [Fe/H] $> -0.5$).  The left panels in Figure \ref{fig:fig4} compare the magnitude distributions of the 377 stars in the higher-latitude combined (m0m75 + m0m10) fields with the 290 stars in the lower-latitude m0m5  field. The split red clump is very evident in the higher latitude sample, and appears to be absent in the m0m5 sample.

In order to interpret these observations, we compare them to an evolutionary N-body model of a boxy/peanut bulge from Athanassoula (2003) (A03). This model started as an isolated, axisymmetric
galaxy with live disk and halo components, such that the disk is
maximum (Sackett 1997; Gerhard 2006). The halo has a volume density described by:

\begin{equation}
\rho_h (r) = \frac {M_h}{2\pi^{3/2}}~~ \frac{\alpha}{r_c} ~~\frac {exp(-r^2/r_c^2)}{r^2+\gamma^2},
\nonumber\\
\end{equation}

\noindent
where $r$ is the radius, $M_h$ is the halo mass, $\gamma$
and $r_c$ are the halo core and cut-off radii, respectively, and the
constant $\alpha$ is given by $\alpha = [1 - \sqrt \pi~~exp (q^2)~~(1
  -erf (q))]^{-1}$, where $q=\gamma / r_c$ (Hernquist 1993). 

The initial density distribution of the disc, for which there are 200 000 particles, is

\begin{equation}
\rho_d (R, z) = \frac {M_d}{4 \pi h^2 z_0}~~exp (- R/h)~~sech^2 (\frac{z}{z_0}),
\end{equation}

where $R$ is the cylindrical radius, $M_d$ is the disc mass, $h$ is the
disc radial scale length and $z_0$ is the disc vertical scale thickness.

We adopted  $z_0/h = 0.2$, r$_c$/h = 10, $M_h/M_d = 5$ and $\gamma$/h = $5$. The velocity dispersion in the disk was chosen such that Toomre's parameter is equal to 1.2 (see A03).

In this model, the bar which grows from the disk, buckles to form a boxy/peanut-bar/bulge (Athanassoula 2005).
Although the model was not designed to match the Galactic bulge, it gives a good representation of its structure and kinematics. 

The right panels in Figure \ref{fig:fig4} show the expected red clump magnitude distribution in the same fields, as derived from this model. The selection criteria for stars taken from the model were identical to that of the observations. To select the roughly constant number of stars per unit magnitude in the data, we scaled the model in $(x,y,z)$ space by matching its projected exponential scaleheight along the minor axis to the observed scaleheight of the Galactic bulge.  In velocity space, the model is scaled to match 
the velocity dispersion of the bulge stars in the ARGOS m0m5 field. The model
is then viewed from the Galactic plane at a distance of $8.0$ kpc from
the center and located so that the observer-center line is at
$20^\circ$ to the major axis of the bar.  The absolute magnitude of the
red clump stars in the model was set so that the minimum of the split red clump
magnitude distribution in the (m0m75 + m0m10) fields of the model is
at $K_\circ = 13.0$, as observed for the ARGOS data in the same
fields. The corresponding absolute magnitude of the red clump is $M_K =
-1.51$.  For comparison, the Alves (2000) calibration is $-1.65$. For the model, the stars have been selected to mimic the roughly constant stars per magnitude in the data. 

The observed and model magnitude distributions are qualitatively
similar. The model magnitude distribution shows clear bimodality in
the higher latitude fields (bottom right panel). In the lower m0m5
field, the bimodality is almost absent (upper right panel).

\begin{figure*}
\centering
\includegraphics[scale=0.36]{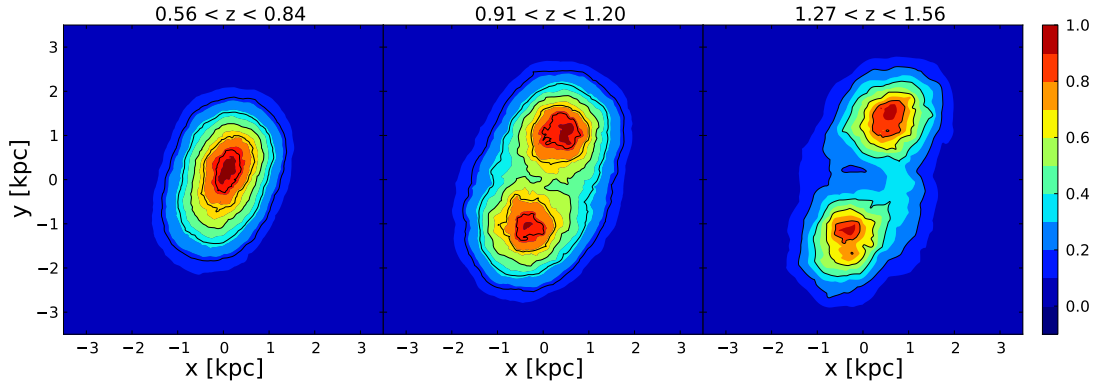} 
\caption{Contour plots of the $(x,y)$ density distribution of the 
N-body model for three slices in $z$: $0.56 < z < 0.84$, $0.91 < z < 1.20$
and $1.27 < z < 1.56$ kpc corresponding to the three ARGOS bulge
fields on the minor axis of the bulge. The sun lies at y = --8 kpc and the angle between the Sun-center line and the major axis of the bar is 20$^\circ$. The higher slices show the dip near the minor axis in the density distribution associated with the structure of the bulge. The colorbar shows the normalised number of stars. This figure illustrates the peanut nature of the bulge model; it is not intended for direct comparison with the data.}
\label{fig:fig5}
\end{figure*}

The orbit structure in the boxy/peanut bulge appears to generate a dip in the density distribution near the minor axis at higher latitudes which produces the split structure in the red clump magnitude distribution. A related model by Shen et al. (2010) is similar. 
We have also examined several of Athanassoula's evolutionary N-body bar/bulge models, run with different initial conditions, and find that the split structure at higher latitudes near the minor axis appears to be a generic feature of bulges formed through disk instability. The detailed structure of the split (e.g. the magnitude separation of the two peaks) varies from model to model.

We now visualize the split structure of the model in Figure \ref{fig:fig5},
which shows the magnitude distributions in three slices of height above
the Galactic plane. 
In the figure, $(x,y)$ are coordinates projected on to
the Galactic plane.  The origin is at the Galactic center, the $y-$axis lies along the Sun-center line, $x$ increases towards decreasing longitude and $z$ is the height above the plane.   This figure is intended to illustrate the structure of the boxy/peanut bulge. It is not intended for direct comparison with the data. It shows
the density distribution in slices of the z-height across the model parallel to the $(x,y)$ plane. The three slices in $z$ are $0.56 < z < 0.84$, $0.91 < z < 1.20$ and $1.27 < z < 1.56$ kpc.
These slices correspond to the range of $z$ for the three ARGOS fields
on the minor axis (m0m5, m0m75 and m0m10) where the line of sight
intersects the $y = 0$ plane.  The figure shows only stars with $|y| < 3.5$ kpc. The highest slice $1.27 < z < 1.56$ kpc shows a decrease in the
density near the minor axis which we can associate with the boxy/peanut
structure of the bulge. The intermediate slice $0.91 < z < 1.20$ kpc also shows the two
density maxima away from the minor axis. The lowest $0.56 < z < 0.84$ kpc slice shows a 
central density peak with almost no sign of the double-peaked structure. This is consistent with the split clump results along the minor axis for the corresponding latitudes of the K magnitude distributions (Figure \ref{fig:fig4}) where no split is seen at m0m5 and the split is seen at m0m75 + m0m10,.
The orbit structure of the boxy/peanut bulge appears to generate a dip in the 
density distribution near the minor axis at higher latitudes, which 
produces the split structure in the red clump magnitude distribution. In projection, 
it would show the weak X-structure observed in other systems: 
e.g. IC 4767 (Whitmore \& Bell 1988).

In the Galactic bulge, it appears that this peanut structure involves only stars with [Fe/H] $>$ --0.5.  We conclude that the disk from which the bulge formed had relatively few stars with [Fe/H] $<$ --0.5. This limit is unlikely to provide a useful guide on the time at which the bulge formation event occurred, because the chemical and dynamical evolution in the inner Galaxy were probably very rapid. 

The model seen in Figure \ref{fig:fig5} indicates that the split in the red clump should widen as the height above the plane increases. Figure \ref{fig:fig5} shows an increasing and more prominent separation between the two peaks in the density distribution with increasing height above the plane. For the stars with [Fe/H] $>-0.5$, which are involved in the split clump, we might therefore expect to see a difference in the K magnitude distribution at m0m75 compared to m0m10.
 
\begin{figure}[!h]
\centering
\includegraphics[scale=0.25]{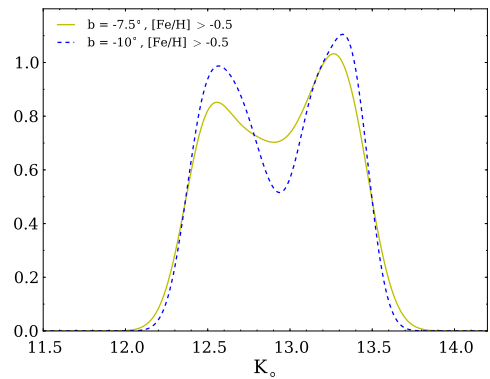} 
\caption{Magnitude distributions for the clump stars with [Fe/H] $>$ --0.5 in m0m75 and m0m10 selected with $12.38 <$ $K_\circ$ $< 13.48$, 1.9 $<$ log g $<$ 3.1. For m0m75 there are 143 stars and for m0m10 there are 147 stars. This figure shows that the clump separation is more clearly defined and is somewhat larger in K-magnitude for the higher latitude field.}
\label{fig:7510}
\end{figure}

Figure \ref{fig:7510} compares the K magnitude distribution for the stars with [Fe/H] $> -0.5$ for the field m0m75 (143 stars) and the field m0m10 (147 stars). The K-magnitude distribution for m0m75 is clearly flatter than for m0m10.  At the higher latitude of m0m10, the K magnitude of the fainter clump is fainter by about $0.06$ mag leading to a slightly wider separation of the peaks for m0m10.  This small increase in separation corresponds to a linear increase in separation of 225 pc. When comparing Figures \ref{fig:fig5} and \ref{fig:7510}, please note that the coordinates in Figure \ref{fig:fig5} are linear, while the horizontal axis of Figure \ref{fig:7510} is in magnitudes.

Figure \ref{fig:fig5} demonstrates the structure of boxy/peanut bulges in the central region.  The flatter K-magnitude distribution for m0m75 and the deeper minimum and slightly wider separation of the peaks at m0m10 seen in our data appear consistent with the structure seen in the 
$|z|-$slices shown in Figure \ref{fig:fig5}.  For the data, taken in slices of latitude rather than height below the plane, the distance from the plane increases with distance along the line of sight, and the relative density distributions of near and far clump stars along the line of sight are therefore affected. We investigate this below.

The ARGOS data are not well suited to investigating the relative numbers of brighter and fainter clump giants, because our stars were selected to be approximately uniform in their K-magnitude distribution. We do not use the ARGOS data to test the decrease in density distribution of the total population of faint red clump compared to bright red clump along the line of sight as a function of latitude reported by Saito et al. (2011); McWilliam \& Zoccali (2010) and Nataf et al. (2010). We do see a bias in the apparent [Fe/H] distribution, with more metal-rich stars located in the near clump (see Figure 3) and more metal-poor stars at fainter magnitudes. This bias is associated with the metallicity gradient in the bulge (see Zoccali (2010), Ness \& Freeman (2012)). The stars on the near side of the split clump are closer to the plane and are therefore more metal-rich in the mean, while the stars on the far side of the clump are at larger $z-$heights and are in the mean more metal-poor. 

\begin{figure*}
\includegraphics[scale=0.36]{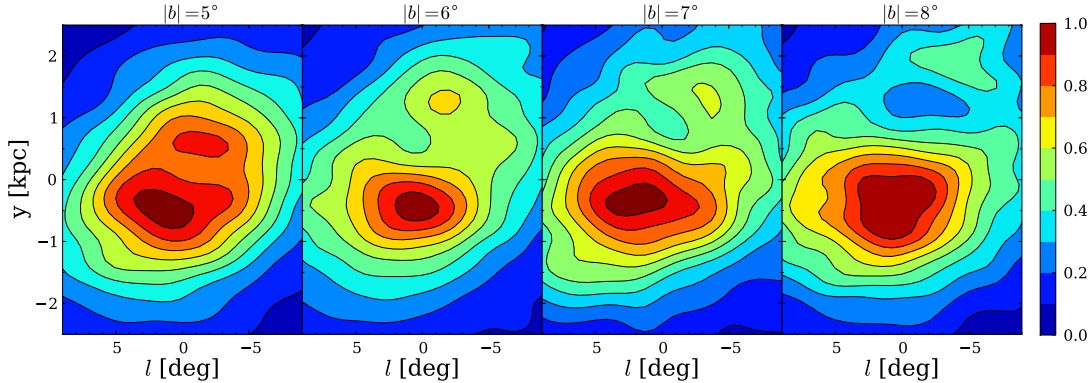}
\caption{The (longitude,$y$) density distribution for the N-body model in four one-degree-wide slices in latitude, at $|b|$ = $5^\circ$, $6^\circ$ $7^\circ$ and $8^\circ$.  The sun lies at $y = -8$ kpc and the angle between the Sun-center line and the major axis of the bar is 20$^\circ$. With increasing distance from the plane, the density of the more distant clump decreases relative to the nearer clump, and the separation between the nearer and more distant clumps increases.  The colorbar shows the normalised number of stars.}
\label{fig:fig7}
\end{figure*}

The photometric studies by Saito et al. (2011), McWilliam \& Zoccali (2010) and Nataf et al. (2010) examine the total density of the red clump stars in the central region.  For the split red clump, these studies show that the near clump dominates and becomes brighter with respect to the faint clump at latitudes further from the plane. We can test for this effect in our model by examining its stellar density distribution in slices of latitude rather than height. Figure \ref{fig:fig7} shows the density distribution of the stars in our model in the inner region in one-degree slices of latitude at $|b| =5^\circ, 6^\circ, 7^\circ$ and $8^\circ$. Our model is symmetrical in $z$ so we combine negative and positive latitudes. These one-degree slices are a subset of those mapped by Saito et al. (2011) at negative and positive latitudes, and are intended for comparison with the Saito et al. (2011) maps. As in the Saito et al. (2011)
data, the number of stars in the far clump in the models decreases relative to the near clump with increasing $|$latitude$|$, and the separation of the two clumps increases.  Along a given line of sight, increasing distance corresponds to larger heights from the plane; at a latitude of $|b| = 8^\circ$, the faint clump stars are about 300 pc further from the plane than the bright clump stars and contribute a significantly smaller fraction of the stellar density than the near red clump. This effect is a consequence of the decreasing density of each arm of the split clump as a function of distance from the plane.

\section{THE KINEMATICS OF THE SPLIT RED CLUMP}

We would like to know whether the stars in the near and far groups
of the magnitude distribution show different kinematic properties
and, if so, whether these are reproduced by the model.
First we examine the radial (line of sight) velocity distributions 
of clump stars in the two latitude samples, m0m5 and the combined m0m75 + m0m10 sample
and compare them with the kinematics of the model. The minimum in the
magnitude distribution of the clump stars is seen at $K = 13.0$, so
we divide the stars into a closer bright group ($12.38 < K < 13.00$) and 
a more distant faint group ($13.00 < K < 13.48$).

\begin{figure}[h]
\includegraphics[scale=0.3]{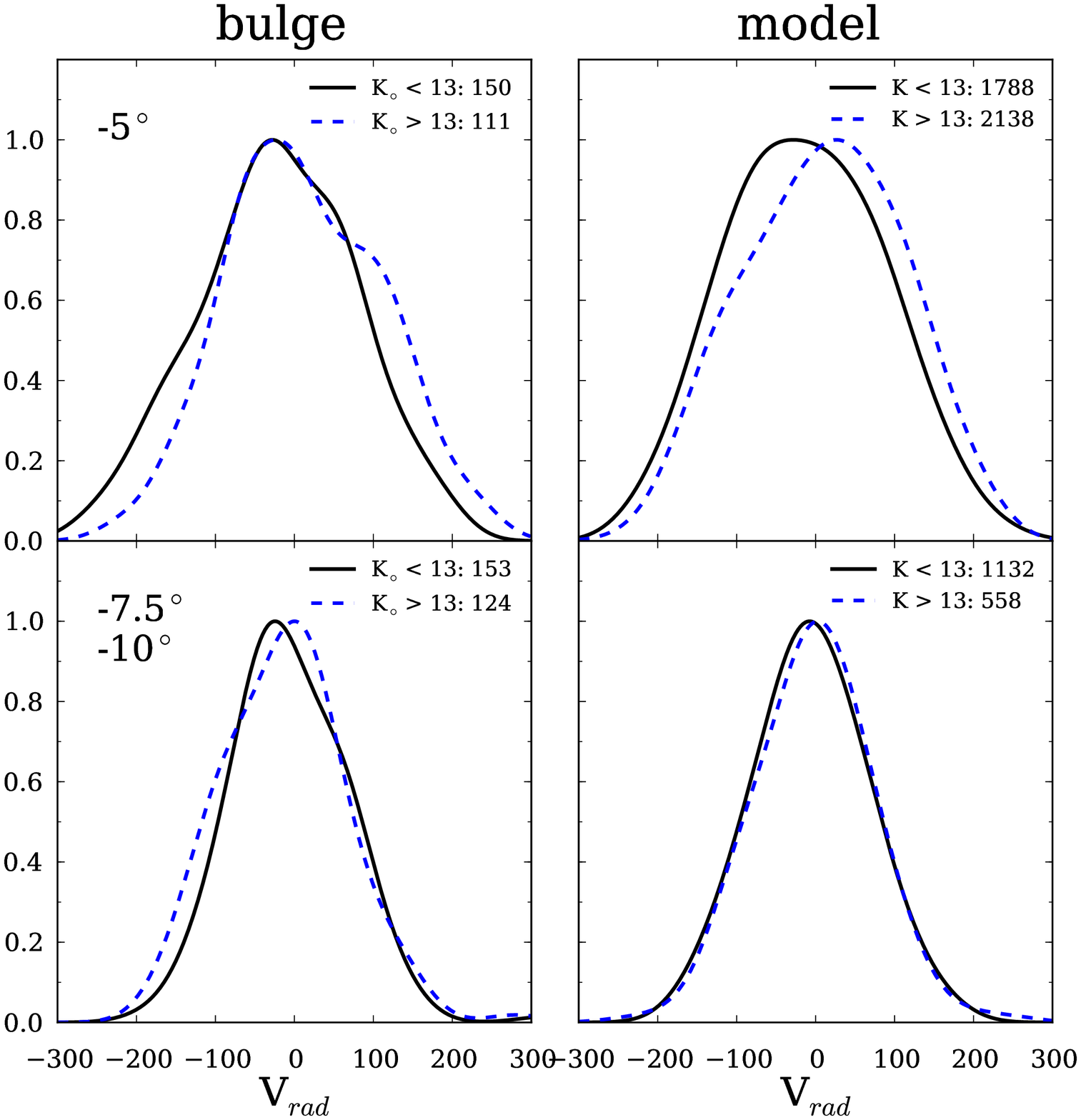} \\
 \caption{Velocity distributions of clump stars selected with $K_\circ$ = 12.38 - 13.48. The central peak is stronger for the higher latitude and split clump, shown as generalised histograms using a Gaussian kernel
with $\sigma = 35$ \kms. {\it Upper left}: the velocity distribution 
for the m0m5 field. The distributions for the near and far groups of
the split clump are shown in the filled and dashed lines respectively. The numbers in
the upper right corner of each panel show the number of stars in each 
group. {\it Lower left}: the velocity distribution for the two combined 
higher-latitude fields. The panels on the right show the velocity 
distributions of the same fields for the model. The normalised units on the vertical axis are arbitrary.}
\label{fig:fig8}
\end{figure}

The left panels of Figure \ref{fig:fig8} show the velocity distributions
for the bright (solid curve) and faint (broken curve) groups in the two observed
latitude samples. The velocity
distributions are presented as generalised histograms, in which
each stellar velocity is represented by a Gaussian with $\sigma =
35$ \kms\ centered at its observed velocity; this avoids binning 
and suppresses high frequency noise in the distribution. 
This representation does not affect the mean velocities given below, which are calculated directly from the data: it is intended to make the form of the distribution more clear visually. The numbers of stars in each subsample are shown in the panels. The right panels show the corresponding velocity
distributions for the model. We recall that the model velocities were scaled
so that the velocity dispersion of the model for the m0m5 sample
agrees with the observed dispersion ($97 \pm 4$ \kms).  The velocity
dispersion of the high latitude sample ($77 \pm 3$ \kms) is also similar
to that of the model ($70 \pm 2$ \kms), with no significant difference
in dispersion between the bright and faint groups. For the low latitude field, the velocity offsets between bright and faint groups are $-30 \pm 12 $ \kms\ for the data and $-24 \pm 3$ \kms\ for the model. For the high latitude fields, the offsets are $ +7 \pm 9$ \kms\  for the data and $ -3 \pm 5$ \kms\ for the model. The conclusion from this figure is that the velocity offsets between the near and far groups of the split clump are similar within the uncertainties for the data and model in both of the latitude regions. These velocity offsets come mainly from streaming motions within the bar/bulge. The velocity offsets in the model are sensitive to the bar orientation relative to the Sun-center line; the velocity offsets agree with the observed offsets, within the errors, for a bar angle of $20^\circ$. 

In this section, we have considered only the kinematics of stars in ARGOS fields on the minor axis. A detailed comparison of the kinematics for the model and the data for the other ARGOS fields, as shown in Figure 1, will be presented in a separate paper (Ness et al. 2012, in preparation). 

\section{CONCLUSION}

The ARGOS survey provides a large sample of bulge stars with accurate 
abundances and radial velocities, relatively free of contamination from foreground and background stars. In this paper we used ARGOS stars near 
the minor axis of the bulge to investigate the properties of the split
red clump.  The split in the red clump magnitude distribution is seen very clearly
in the higher latitude fields ($b = -6.5^\circ$ to $-11^\circ$) near the 
minor axis, for stars with [Fe/H] $> -0.5$. The more metal-poor red clump stars ([Fe/H]$ < -0.5$) in the same regions of the bulge do not show the split, and it is almost absent at all metallicities 
for our m0m5 field ($b = -4^\circ$ to $-6^\circ$).

We compared the structure and kinematics of the ARGOS stars with an
N-body model of a bulge. In the model, the boxy/peanut bulge grew 
through the bar-forming and bar-buckling instability of a disk of stars. Although the model was not constructed to fit the Galactic bulge, it is a good representation. The orbital structure of the boxy/peanut bulge leads to the central depression in its peanut-shaped density distribution: see Athanassoula (2005). The central depression in turn leads to a split bimodal structure along the line of sight at higher latitudes which closely resembles the split observed in the ARGOS stars and in the Saito et al. (2011) data. At low latitude, the kinematics of the bulge stars in the bright and faint groups of the split red clump are also similar to those for the corresponding stars in the model. 

In projection, the density distribution would show an X-shaped structure, particularly after unsharp masking: see Whitmore \& Bell (1988), Athanassoula (2005) and Bureau et al. (2006). It seems likely that the split red clump seen in the Galactic bulge is a generic feature of boxy/peanut bulges which grew from disks. We also conclude that the disk from which the bar/bulge formed would have had relatively few stars with [Fe/H] $< -0.5$.

\section*{ACKNOWLEDGEMENTS}

We thank the Anglo-Australian Observatory, who have made this project possible.

This publication makes use of data products from the Two Micron All Sky Survey, which is a joint project of the University of Massachusetts and the Infrared Processing and Analysis Center/California Institute of Technology, funded by the National Aeronautics and Space Administration and the National Science Foundation.

This work has been supported by the RSAA and Australian Research Council grant DP0988751. 
EA gratefully acknowledges financial support by the European Commission through the DAGAL Network (PITN-GA-2011-289313). J.BH is supported by an ARC Federation Fellowship. G.F.L thanks the Australian research council for support through his Future Fellowship (FT100100268) and Discovery Project (DP110100678). R.R.L. gratefully acknowledges support from the Chilean {\sl Centro de Astrof\'\i sica} FONDAP No. 15010003. L.L.K is supported by the Lend\"ulet program of the Hungarrian Academy of Sciences and the Hungarian OTKA Grants K76816, MB08C 81013 and K83790.


\begin{thebibliography}{28}
\expandafter\ifx\csname natexlab\endcsname\relax\def\natexlab#1{#1}\fi

\bibitem[{{Alves}(2000)}]{Alves2000}
{Alves}, D.~R. 2000, ApJ, 539, 732

\bibitem[{{Athanassoula}(2003)}]{Athanassoula2003}
{Athanassoula}, E. 2003, MNRAS, 341, 1179 (A03)

\bibitem[{{Athanassoula}(2005)}]{Athanassoula2005}
---. 2005, MNRAS, 358, 1477

\bibitem[{{Bessell} \& {Brett}(1988)}]{Bessell1988}
{Bessell}, M.~S. \& {Brett}, J.~M. 1988, PASP, 100, 1134

\bibitem[{{Bureau} {et~al.}(2006){Bureau}, {Aronica}, {Athanassoula},
  {et~al.}}]{Bureau2006}
{Bureau}, M., {Aronica}, G., {Athanassoula}, E., {et~al.} 2006, MNRAS, 370, 753

\bibitem[{{Cassisi} {et~al.}(2006){Cassisi}, {Pietrinferni}, {Salaris},
  {et~al.}}]{Basti2006}
{Cassisi}, S., {Pietrinferni}, A., {Salaris}, M., {et~al.} 2006, MEMSAI, 77, 71

\bibitem[{{Castelli} \& {Kurucz}(2003)}]{CastelliKurucz2003}
{Castelli}, F. \& {Kurucz}, R.~L. 2003, in IAU Symposium, Vol. 210, Modelling
  of Stellar Atmospheres, ed. {N.~Piskunov, W.~W.~Weiss, \& D.~F.~Gray}, 20P

\bibitem[{{De Propris} {et~al.}(2011){De Propris}, {Rich}, {Kunder},
  {et~al.}}]{deProp2011}
{De Propris}, R., {Rich}, R.~M., {Kunder}, A., {et~al.} 2011, ApJL, 732, L36

\bibitem[{{Dwek} {et~al.}(1995){Dwek}, {Arendt}, {Hauser}, {et~al.}}]{Dwek1995}
{Dwek}, E., {Arendt}, R.~G., {Hauser}, M.~G., {et~al.} 1995, ApJ, 445, 716

\bibitem[{{Gerhard}(2002)}]{Gerhard2002}
{Gerhard}, O. 2002, in PASP, Vol. 273, The Dynamics, Structure \& History of
  Galaxies: A Workshop in Honour of Professor Ken Freeman, ed. {G.~S.~Da Costa,
  E.~M.~Sadler, \& H.~Jerjen}, 73

\bibitem[{{Gerhard}(2006)}]{Gerhard2006}
{Gerhard}, O. 2006, in EAS Publications Series, Vol.~20, EAS Publications
  Series, ed. G.~A. {Mamon}, F.~{Combes}, C.~{Deffayet}, \& B.~{Fort}, 89--96

\bibitem[{{Hernquist}(1993)}]{Hernquist1993}
{Hernquist}, L. 1993, ApJS, 86, 389

\bibitem[{{Hinkle} {et~al.}(2000){Hinkle}, {Wallace}, {Harmer}, {Ayres}, \&
  {Valenti}}]{Hinkle2000}
{Hinkle}, K., {Wallace}, L., {Harmer}, D., {Ayres}, T., \& {Valenti}, J. 2000,
  in IAU Joint Discussion, Vol.~1, IAU Joint Discussion

\bibitem[{{Kerr} \& {Lynden-Bell}(1986)}]{KerrLB86}
{Kerr}, F.~J. \& {Lynden-Bell}, D. 1986, MNRAS, 221, 1023

\bibitem[{{L{\'o}pez-Corredoira} {et~al.}(2005){L{\'o}pez-Corredoira},
  {Cabrera-Lavers}, \& {Gerhard}}]{Lopez2005}
{L{\'o}pez-Corredoira}, M., {Cabrera-Lavers}, A., \& {Gerhard}, O.~E. 2005,
  AAP, 439, 107

\bibitem[{{McWilliam} \& {Zoccali}(2010)}]{McWilliamZoccali2010}
{McWilliam}, A. \& {Zoccali}, M. 2010, ApJ, 724, 1491

\bibitem[{{Nataf} {et~al.}(2010){Nataf}, {Udalski}, {Gould}, {Fouqu{\'e}}, \&
  {Stanek}}]{Nataf2010}
{Nataf}, D.~M., {Udalski}, A., {Gould}, A., {Fouqu{\'e}}, P., \& {Stanek},
  K.~Z. 2010, ApJL, 721, L28

\bibitem[{{Ness} \& {Freeman}(2012)}]{Ness2012b}
{Ness}, M. \& {Freeman}, K. 2012, in European Physical Journal Web of
  Conferences, Vol.~19, European Physical Journal Web of Conferences, 6003

\bibitem[{{Rangwala} \& {Williams}(2009)}]{Rang2009}
{Rangwala}, N. \& {Williams}, T.~B. 2009, ApJ, 702, 414

\bibitem[{{Sackett}(1997)}]{Sackett1997}
{Sackett}, P.~D. 1997, ApJ, 483, 103

\bibitem[{{Saito} {et~al.}(2011){Saito}, {Zoccali}, {McWilliam},
  {et~al.}}]{Saito2011}
{Saito}, R.~K., {Zoccali}, M., {McWilliam}, A., {et~al.} 2011, aj, 142, 76

\bibitem[{{Schlegel} {et~al.}(1998){Schlegel}, {Finkbeiner}, \&
  {Davis}}]{Schlegel1998}
{Schlegel}, D.~J., {Finkbeiner}, D.~P., \& {Davis}, M. 1998, ApJ, 500, 525

\bibitem[{{Sharp} {et~al.}(2006){Sharp}, {Saunders}, {Smith},
  {et~al.}}]{Sharp2006}
{Sharp}, R., {Saunders}, W., {Smith}, G., {et~al.} 2006, in SPIE Conference
  Series, Vol. 6269, SPIE Conference Series

\bibitem[{{Shen} {et~al.}(2010){Shen}, {Rich}, {Kormendy}, {et~al.}}]{Shen2010}
{Shen}, J., {Rich}, R.~M., {Kormendy}, J., {et~al.} 2010, ApJL, 720, L72

\bibitem[{{Sneden}(1973)}]{Sneden1973}
{Sneden}, C.~A. 1973, PhD thesis, THE UNIVERSITY OF TEXAS AT AUSTIN.

\bibitem[{{Whitmore} \& {Bell}(1988)}]{WhitBell}
{Whitmore}, B.~C. \& {Bell}, M. 1988, ApJ, 324, 741

\bibitem[{{Zhao} {et~al.}(2001){Zhao}, {Qiu}, \& {Mao}}]{Zhao2001}
{Zhao}, G., {Qiu}, H.~M., \& {Mao}, S. 2001, ApJL, 551, L85

\bibitem[{{Zoccali}(2010)}]{Zoccali2010}
{Zoccali}, M. 2010, in IAU Symposium, Vol. 265, IAU Symposium, ed. K.~{Cunha},
  M.~{Spite}, \& B.~{Barbuy}, 271--278

\end{thebibliography}

\appendix

\section{APPENDIX}

\begin{figure}[h]
\centering
\includegraphics[scale=0.25]{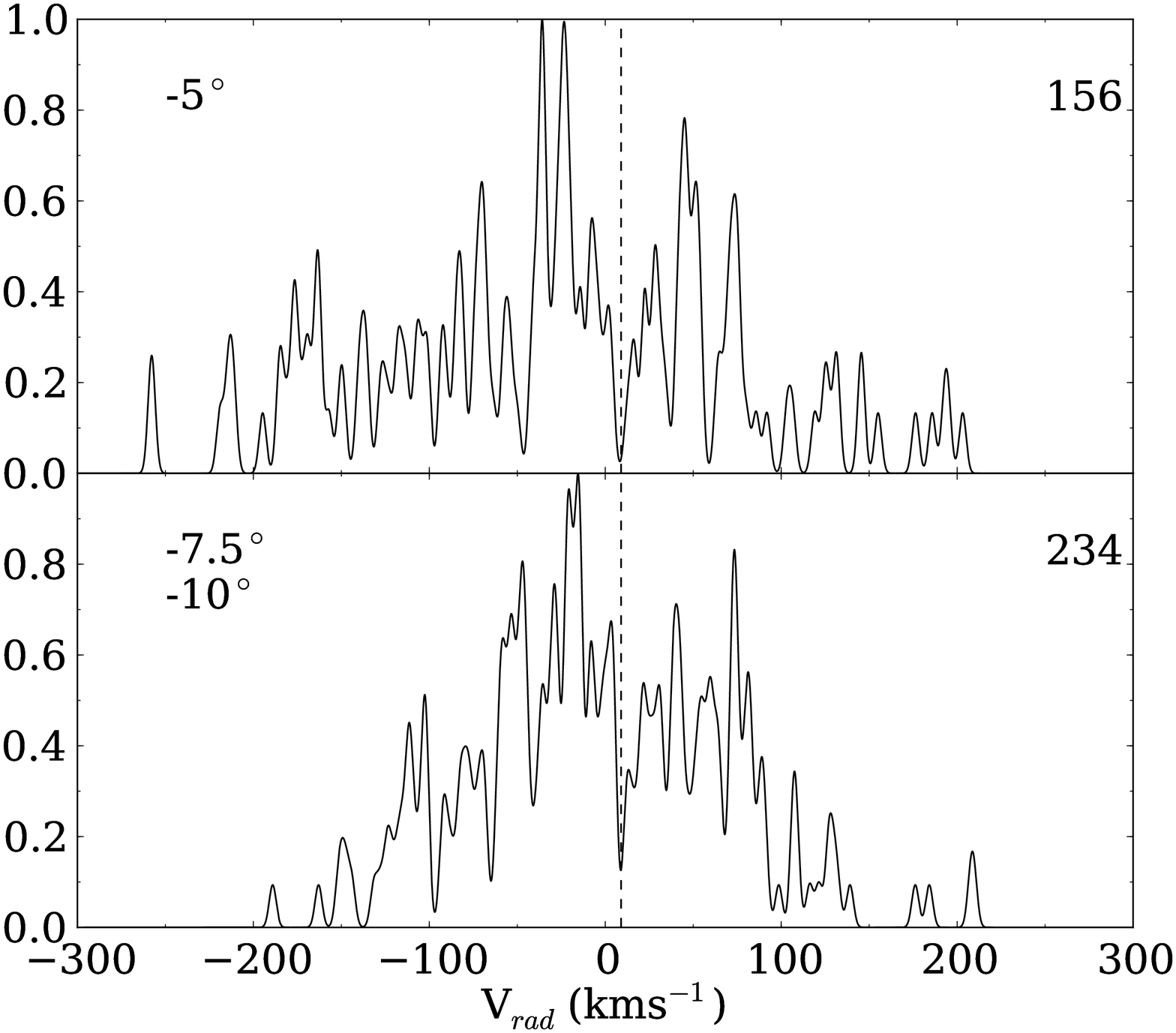}
\caption{Velocity distributions for stars in the bright group of
the split clump, shown as generalised histograms using a Gaussian
kernel with $\sigma = 2$ \kms.  {\it Upper panel}: velocity distribution 
for the m0m5 field. {\it Lower panel}: velocity distribution for the
combined m0m75 and m0m10 fields. The numbers of stars in each subsample
are shown in the upper right corner of each panel.  The vertical dotted 
line at $V_{rad} = 9$ \kms\ shows the location of an apparent deep and
narrow dip in the velocity distribution for both subsamples. No
corresponding feature is seen for the stars in the faint group or in the
model. The normalised units on the vertical axis are arbitrary.}
\label{fig:fig9}
\end{figure}

An unexpected feature is seen in the velocity distribution of the
bright group of clump stars on the minor axis of the bulge. Figure \ref{fig:fig9} compares the generalised 
histograms for the Galactocentric velocity distributions of stars in the bright group
for the m0m5 field and for the higher latitude m0m75 and m0m10 sample.  
In this figure we use a much narrower kernel for the generalised histogram
($\sigma = 2$
\kms, about double the measuring errors) to see the small scale structure. We can see this structure only because we have small measurement errors on our velocities of $< 1.2$ \kms\ which enables us to use this small kernel. Although the figure appears noisy because of the relatively small number of stars and the narrow kernel, both of the samples show a narrow and deep
aligned minimum at a (heliocentric) radial velocity of $9$ \kms. No corresponding feature is seen in the model. 

To evaluate the probability of finding such an alignment of sharp deep minima by chance, we made a Monte Carlo simulation of pairs of samples. We are not testing whether the two distributions in Figure \ref{fig:fig9} are drawn fromthe same population: their different velocity dispersions show that they are not.  We are estimating the probability that two random samples like those in the figure could have such deep aligned minima.  In the simulation, the underlying distributions were gaussian with the same velocity dispersions as the two samples shown in Figure \ref{fig:fig9}.  We drew 1000 samples of 156 and 234 stars respectively from the two samples and found that aligned minima as deep as those seen in Figure \ref{fig:fig9} occurred with a probability of only 0.007. Nevertheless
this feature may not be real, and we draw attention to its presence without 
offering any suggestions for how it might come about.

\end{document}